\documentclass{article}
\usepackage{color}
\usepackage{spconf,amsmath,graphicx,amssymb,delarray,subfigure,verbatim,booktabs}
\usepackage{diagbox, makecell}
\usepackage{url}
\usepackage{hyperref}
\usepackage[numbers,compress]{natbib}
\usepackage{multirow}
\usepackage{bbding}
\usepackage[english]{babel}
\usepackage{lipsum}
\usepackage[utf8]{inputenc}
\usepackage[nolist]{acronym}
\usepackage{pgfplots}
\usepackage[subtle]{savetrees}  
\pgfplotsset{compat=1.15}
\usepgfplotslibrary{colormaps,groupplots,dateplot}
\usepackage{booktabs}

\let\OLDthebibliography\thebibliography
\renewcommand\thebibliography[1]{
	\OLDthebibliography{#1}
	\setlength{\parskip}{0pt}
	\setlength{\itemsep}{1.5pt plus 2ex}
}
\hypersetup{hidelinks,
	colorlinks=true,
	allcolors=black,
	pdfstartview=Fit,
	breaklinks=true}

\title{TaylorBeamixer: Learning Taylor-Inspired All-Neural Multi-Channel Speech Enhancement from Beam-Space Dictionary Perspective}
\name{Andong Li$^{\star \dagger}$, Guochen Yu$^{\star}$, Wenzhe Liu$^{\ddagger}$, Xiaodong Li$^{\star \dagger}$, Chengshi Zheng$^{\star \dagger}$}
\address{$^{\star}$ Key Laboratory of Noise and Vibration Research, Institute of Acoustics, Chinese Academy\\
	of Sciences, Beijing, China\\
	$^{\dagger}$ University of Chinese Academy of Sciences, Beijing, China\\
$^{\ddagger}$ Tencent Ethereal Audio Lab, Tencent Corporation, Shenzhen, China}
\begin{document}
\ninept
\maketitle
\newcommand\blfootnote[1]{%
	\begingroup
	\renewcommand\thefootnote{}\footnote{#1}%
	\addtocounter{footnote}{-1}%
	\endgroup
}

\begin{abstract}
\vspace{-0.2cm}
Despite the promising performance of existing frame-wise all-neural beamformers in the speech enhancement field, it remains unclear what the underlying mechanism exists. In this paper, we revisit the beamforming behavior from the beam-space dictionary perspective and formulate it into the learning and mixing of different beam-space components. Based on that, we propose an all-neural beamformer called TaylorBM to simulate Taylor's series expansion operation in which the 0th-order term serves as a spatial filter to conduct the beam mixing, and several high-order terms are tasked with residual noise cancellation for post-processing. The whole system is devised to work in an end-to-end manner. Experiments are conducted on the spatialized LibriSpeech corpus and results show that the proposed approach outperforms existing advanced baselines in terms of evaluation metrics.

\end{abstract}
\vspace{-0.15cm}
\section{Introduction}
\label{sec:intro}
\vspace{-0.15cm}
By virtue of the spatial information, multi-channel speech enhancement (MC-SE) can effectively extract the target speech from the noisy mixture and often leads to superior performance over the single-channel (SC) case~{\cite{gannot2017consolidated, doclo2015multichannel}}, rendering it convenient for people to stay in connection by remote communication systems during the COVID-19 pandemic period.

Recently, with the advent of deep neural networks (DNNs), we have witnessed the proliferation of neural beamformers (NBFs) by leaps and bounds, which make significant progress over traditional spatial filters~{\cite{heymann2016neural, heymann2015blstm, wang2020complex, ochiai2017unified}}. Existing methods can be broadly broken into three categories. The first class works in a hybrid mode, \emph{i.e.}, the speech/noise mask is estimated by a general network, and a traditional utterance- or batch-level beamformer is utilized for spatial filtering~{\cite{heymann2016neural, heymann2015blstm}}. A critic is that the two modules are often separately tackled, the performance is often limited and suffers from heavy degradation when the frame-wise processing is required. The second one follows the \textit{extraction-fusion} protocol where the spectral and spatial cues are explicitly/implicitly extracted, and the network serves as the fusion module to combine both features in the non-linear space to derive the target speech in an end-to-end (E2E) manner~{\cite{wang2020complex, gu2020enhancing, chakrabarty2019time, tan2022neural, tesch2022insights}}. As a natural extension of SC-SE, they often cause non-linear speech distortion because the spatial discrimination property is not fully utilized. For the third class, more recently, a few studies reveal the potential and superiority of frame-wise all-neural beamformers in either time-domain~{\cite{luo2020end}} or time-frequency (T-F) domain~{\cite{xiao2016deep, zhang2021adl, li2022embedding}}, where DNNs are employed to replace or abstract part of the signal-processing based operations to estimate the beamforming weights. As the beamforming weights are non-linearly mapped frame by frame, less algorithmic delay is required while the performance can be guaranteed under the E2E training criterion.

To enable frame-wise all-neural beamformers, the spatial and spectral modes are usually entangled in the non-linear feature space, and the whole system is usually encapsulated into a black box, thus lacking adequate interpretability and transparency~{\cite{li2022embedding}}. As the first trial to open up the box, in~{\cite{li2022taylorbeamformer}}, we proposed to decouple the pipeline into the superimposition of the spatial and spectral processing modes based on Taylor's approximation theory, where the 0th-order term corresponds to the spatial filter and the remaining several high-order terms are designed to cancel the residual interference in the spectral sense. Despite the deep insight, we find it still unclear or even unknown \textit{where does the generated spatial beam in the 0th-order term come from?} In other words, \textit{given the noisy complex spectra from the array as the input, the estimated beam seems to be ``created" from scratch via the network without any explicit prior representations}. In this study, we revisit the spatial filtering behavior from the beam-space dictionary perspective, and formulate the beamforming operation into the adaptive activating and mixing of different basis beams in the beam-space domain~{\cite{liu2022neural}}. Based on that, we propose a novel all-neural beamformer called TaylorBeamixer (abbreviated as \textbf{TaylorBM}) based on Taylor's approximation theory. Specifically, the 0th-order term serves as the spatial filter by dynamically selecting and mixing the beam components with different spatial responses. And multiple high-order terms are superimposed as the residual noise canceller for post-processing. To enable the E2E training, we replace the complicated derivative terms with trainable modules. Compared with our preliminary work~{\cite{li2022taylorbeamformer}}, we merely add one differentiable layer with neglectable parameters. Nonetheless, it provides a different and new perspective on the beamforming process and also achieves on-par or better performance. We hope this work can take a further step toward understanding the internal logic of the white-box-oriented NBFs.

The rest of the paper is organized as follows. In Section~{\ref{sec:physical-model}}, we formulate the problem. In Section~{\ref{sec:proposed-approach}}, the proposed method is presented. Section~{\ref{sec:experimental-setup}} gives the experimental setup, and results and analysis are presented in Section~{\ref{sec:results-and-analysis}}. Conclusions are drawn in Section~{\ref{sec:conclusion}}.
\vspace{-0.35cm}
\section{PORBLEM FORMULATION}
\vspace{-0.15cm}
\label{sec:physical-model}
Given a recorded $M$-channel time-domain acoustic signal vector $\mathbf{x}\left(n\right)\in\mathbb{R}^{M}$ in a reverberant and noisy environment, the physical model after the short-time Fourier transform (STFT) can be given:
\begin{equation}
\label{eqn:eq1}
\mathbf{X}_{l, k} = \mathbf{c}_{k}S_{l, k} + \mathbf{V}_{l, k} + \mathbf{N}_{l, k},
\end{equation}
where $\left\{\mathbf{X}_{l, k}, \mathbf{V}_{l, k}, \mathbf{N}_{l, k}\right\}\in\mathbb{C}^{M}$ denote the mixture, reverberation, and noise components in the frequency index of $k\in\left\{1,\cdots,K\right\}$ and time index of $l\in\left\{1,\cdots,L\right\}$. $\mathbf{c}_{k}\in\mathbb{C}^{M}$ is the acoutic transfer function (ATF) vector of speech and $S_{l, k}\in\mathbb{C}$ is the complex spectrum of the clean speech.

To suppress the directional noise and reverberation components, the beamforming technique is commonly adopted, given by
\begin{equation}
\label{eqn:eq2}
\vspace{-0.1cm}
\widetilde{S}_{l, k} = \mathbf{W}_{l, k}^{\text{H}}\mathbf{X}_{l, k},
\vspace{-0.1cm}
\end{equation}
where $\mathbf{W}_{l, k}\in\mathbb{C}^{M}$ denotes the beamforming weights. $\widetilde{\cdot}$ and $\left(\cdot\right)^\text{H}$ denote the estimated variable and Hermitian transpose, respectively.

In~{\cite{pan2022framework}}, the adaptive beamformer was decomposed into the product of a fixed beamformer (FB) and a post filter (PF):
\begin{equation}
\label{eqn:eq3}
\vspace{-0.0cm}
\mathbf{W}_{l, k} = \mathbf{B}_{\text{Fix},k}G_{l, k},
\vspace{-0.0cm}
\end{equation}
where $\mathbf{B}_{\text{Fix}, k}$ is a time-invariant fixed beam and $G_{l, k}$ is a controlling coefficient in each T-F bin. However, as the desired speech source may appear in any spatial position with different direction-of-arrivals (DOAs), it seems far from adequate to track and approximate the adaptive beamformer with merely one set of FB and PF. Towards this end, we revisit the beamforming process and formulate it into the adaptive activating and mixing of a set of beam-space components. Concretely, we define a time-invariant beam-space dictionary (TI-BD) $\mathcal{B}=\left(\mathbf{B}_{1},\cdots,\mathbf{B}_{P}\right)\in\mathbb{C}^{K\times M\times P}$, where $\mathbf{B}_{p}\in\mathbb{C}^{K\times M}$ refers to the basis beam with index $p\in\left\{1,\cdots,P\right\}$. To control the gain, one can allocate each beam  with a different activating coefficient, and the activating matrix can be defined as $\mathcal{G} = \left(\mathbf{G}_{1},\cdots,\mathbf{G}_{P}\right)\in\mathbb{C}^{L\times K\times P}$. Therefore, the decomposition in Eq.({\ref{eqn:eq3}}) is converted into a more generalized format, \emph{i.e.},
\begin{equation}
\label{eqn:eq4}
\vspace{-0.2cm}
\mathbf{W}_{l, k} = \sum_{p=1}^{P}\mathcal{B}_{k,:,p}\mathcal{G}_{l,k,p}.
\vspace{-0.2cm}
\end{equation}

Recall that in the non-negative matrix factorization (NMF) based SE methods~{\cite{bando2018statistical, lee2000algorithms}}, a similar mathematical expression has been given, but they are by no means the same thing. The major difference is that the dictionary herein is built in the beam-space domain while the dictionary in the NMF-based SE is built in the frequency domain. Besides, the non-negative property does not hold herein as the spatial beam is calculated in the complex-valued format.
\begin{figure*}[t]
	\centering
	\centerline{\includegraphics[width=1.60\columnwidth]{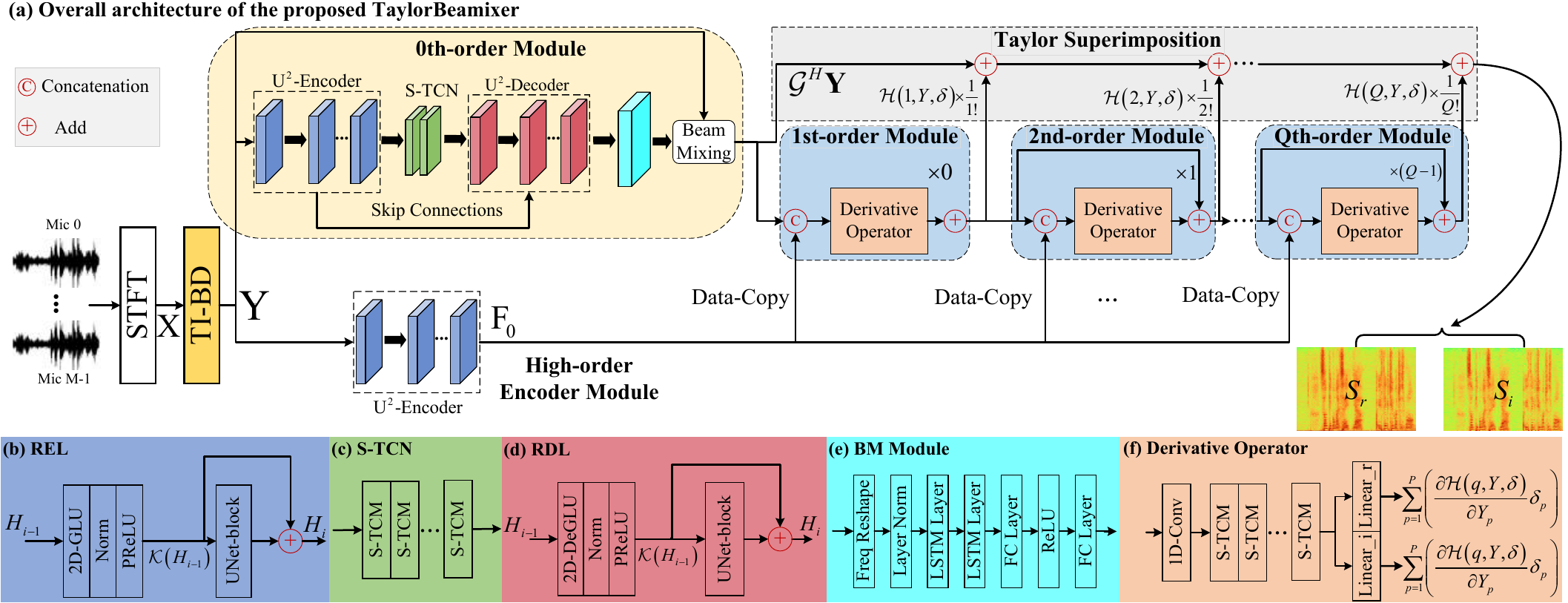}}
	\caption{The diagram of the proposed TaylorBM. Different modules are remarked with different colors.}
	\label{fig:architecture}
	\vspace{-0.4cm}
\end{figure*}
Substituting Eq.({\ref{eqn:eq4}}) into Eq.({\ref{eqn:eq2}}), one can get
\vspace{-0.2cm}
\begin{small}
\begin{equation}
\label{eqn:eq5}
\vspace{-0.1cm}
\widetilde{S}_{l, k} = \sum_{p=1}^{P}\mathcal{G}^{\text{H}}_{l,k,p}Y_{l,k,p},
\vspace{-0.2cm}
\end{equation}
\end{small}where $Y_{l,k,p} \overset{\text{def}}{=} \mathcal{B}^{\text{H}}_{k,:,p}\mathbf{X}_{l,k}\in\mathbb{C}$ refers to the obtained beam output by the $p$th basis beam. As such, we provide a different insight toward the beamforming process, \emph{i.e.}, the beamforming operation can be regarded as the mixing strategy within the beam-space dictionary weighted by different activating coefficients. This is in essence a type of dictionary learning~{\cite{tovsic2011dictionary, mairal2009online}}.

We expect the target speech to be distortionless after the beamforming, \emph{e.g.}, minimum variance distortionless response (MVDR), the beamforming output signal can thus be given by
\begin{equation}
\label{eqn:eq6}
\vspace{-0.1cm}
\widetilde{S}_{l,k} = \sum_{p=1}^{P}\mathcal{G}^{\text{H}}_{l,k,p}Y_{l,k,p} = S_{l,k} + \sum_{p=1}^{P}\mathcal{G}^{\text{H}}_{l,k,p}\mathcal{B}^{\text{H}}_{k,:,p}\mathbf{R}_{l,k},
\vspace{-0.1cm}
\end{equation}
where $\mathbf{R}_{l,k} = \mathbf{V}_{l,k} + \mathbf{N}_{l,k}$. Assuming there exists a prior term $\delta_{l,k,p}$ for each beam, which aims to cancel the residual noise after the summation, one can get
\begin{small}
\begin{equation}
\label{eqn:eq7}
\vspace{-0.1cm}
S_{l,k} = \sum_{p=1}^{P}\mathcal{G}_{l,k,p}^{\text{H}}\left(Y_{l,k,p} + \delta_{l,k,p}\right),
\vspace{-0.1cm}
\end{equation}
\end{small}where $\delta_{l,k,p} = -\mathcal{B}^{\text{H}}_{l,k,p}\mathbf{R}_{l,k}$ will be discussed later. One can see that if the each beam can introduce the prior term and add it in advance, then we can recover the target speech perfectly in theory. From now on, we will drop the subscript $\left\{l, k\right\}$ if no confusion arises. We abstract the operation of weighting each beam as a general function $F_{p}\left(\cdot\right)$, and assume the function to be differentiable to each order, then we can resolve Eq.~(\ref{eqn:eq7}) with infinite Taylor's series expansion at $Y_{p}$ as
\begin{equation}
\label{eqn:eq8}
S = \sum_{p=1}^{P}F_{p}\left(Y_{p}\right) + \sum_{q=1}^{+\infty}\frac{1}{q!}\sum_{p=1}^{P}\frac{\partial^{q}F_{p}\left(Y_{p}\right)}{\partial^{q}Y_{p}}\delta_{p}^q,
\end{equation}
where the 0th-order represents the behavior of spatial filtering and high-order terms serve as the residual noise canceller for post-processing. Note that in~{\cite{li2022taylorbeamformer}}, a similar format was derived. However, in this work, we provide a more intuitive explanation of the beamforming behavior, \emph{i.e.}, a set of beam components are first generated by the time-invariant beam-space dictionary in advance, followed by an adaptive activating matrix to mix and estimate the target spatial beam. In contrast, in~{\cite{li2022taylorbeamformer}}, it remains unclear about the internal mechanism of the time-variant beam generation. 
\vspace{-0.35cm}
\section{Proposed Approach}
\label{sec:proposed-approach}
\vspace{-0.2cm}
\subsection{Relation Between Adjacent Order Terms}
\vspace{-0.2cm}
In practical implementation, we usually truncate the order number into a finite value, \emph{i.e.}, $Q$. To resolve Eq.(\ref{eqn:eq8}), let us notate the $q$th order term as $\mathcal{H}\left(q, Y, \delta\right)$, shown as
\begin{equation}
\label{eqn:eq9}\vspace{-0.1cm}
\mathcal{H}\left(q, Y, \delta\right) = \sum_{p=1}^{P}\frac{\partial^{q}F_{p}\left(Y_{p}\right)}{\partial^{q}Y_{p}}\delta_{p}^q.
\vspace{-0.1cm}
\end{equation}

Note that the factorial term is dropped for convenience. To obtain the relation between adjacent orders, we differentiate $\mathcal{H}\left(q, Y, \delta\right)$ with respect to $Y_{p}$:
\begin{equation}
\label{eqn:eq10}
\vspace{-0.1cm}
\frac{\partial\mathcal{H}\left(q, Y, \delta\right)}{\partial Y_{p}} = \frac{\partial}{\partial Y_{p}}\left(\frac{\partial^{q}F_{p}\left(Y_{p}\right)}{\partial^{q}Y_{p}}\delta_{p}^q\right) + \frac{\partial}{\partial Y_{p}}\left(\sum_{p^{'}\neq p}\frac{\partial^{q}F_{p^{'}}\left(Y_{p^{'}}\right)}{\partial^{q}Y_{p^{'}}}\delta_{p^{'}}^q\right).
\vspace{-0.1cm}
\end{equation}

Ideally, if each acoustic source lies within a different beam component, neighboring beams can be approximately assumed as statistically mutually independent. The more the number of microphones and beams,  the better the independence assumption can hold. Meanwhile, we can control the orthogonality of beams by selecting a suitable beam-space dictionary. To simplify the derivation, we assume the statistical independence between $Y_{p}$ and $Y_{p^{'}}$ for $\forall p^{'}\neq p$. Eq.~(\ref{eqn:eq10}) can then be further converted according to the chain rule: 
\begin{equation}
\label{eqn:eq11}
\vspace{-0.1cm}
\frac{\partial\mathcal{H}\left(q, Y, \delta\right)}{\partial Y_{p}} = \frac{\partial}{\partial Y_{p}}\left(\frac{\partial^{q}F_{p}\left(Y_{p}\right)}{\partial^{q}Y_{p}}\right)\delta_{p}^q + \frac{\partial^{q}F_{p}\left(Y_{p}\right)}{\partial^{q}Y_{p}}\frac{\partial\delta_{p}^q}{\partial Y_{p}}.
\vspace{-0.1cm}
\end{equation}

Notice that,
\begin{gather}
\label{eqn:eq12}
\vspace{-0.1cm}
\sum_{p=1}^{P}\frac{\partial}{\partial Y_{p}}\left(\frac{\partial^{q}F_{p}\left(Y_{p}\right)}{\partial^{q}Y_{p}}\right)\delta^{(q+1)}_{p} = \mathcal{H}\left(q+1, Y, \delta\right),\\
\sum_{p=1}^{P}\frac{\partial^{q}F_{p}\left(Y_{p}\right)}{\partial^{q}Y_{p}}\frac{\partial\delta_{p}^q}{\partial Y_{p}}\delta_{p} = -q\sum_{p=1}^{P}\frac{\partial^{q}F_{p}\left(Y_{p}\right)}{\partial^{q}Y_{p}}\delta^{q}_{p} = -q\mathcal{H}\left(q, Y, \delta\right).
\vspace{-0.1cm}
\end{gather}

We can thus derive the recursive formula between $\mathcal{H}\left(q, Y, \delta\right)$ and $\mathcal{H}\left(q+1, Y, \delta\right)$ as
\begin{equation}
\label{eqn:eq13}
\mathcal{H}\left(q+1, Y, \delta\right) = q\mathcal{H}\left(q, Y, \delta\right) + \sum_{p=1}^{P}\frac{\partial\mathcal{H}\left(q, Y, \delta\right)}{\partial Y_{p}}\delta_{p}.
\vspace{-0.1cm}
\end{equation}

One can get that there exists one term in the right-hand of Eq.~(\ref{eqn:eq13}) that involves both the derivative operation and $\delta_{p}$. Moreover, we actually do not know its real distribution. Similar to~{\cite{li2022taylorbeamformer, li2022taylor}}, we replace the complicated term with a trainable network module and learn it directly from training data automatically. Besides, as the derivative operation is avoided, the training process can thus be more stable.
\vspace{-0.15cm}
\subsection{Time-invariant Beam-space Dictionary}
\label{sec:beam-dictionary}
\vspace{-0.15cm}
To analyze the impact of the beam-space dictionary, three TI-BD tactics are investigated, namely fixed, semi-learnable, and full-learnable. For fixed type, we select two classical FBs as the candidate, namely delay-and-sum (DS) and superdirective (SD)~{\cite{benesty2008microphone}}, which can be expressed:
\vspace{-0.1cm}
\begin{equation}
\label{eqn:eq14}
\vspace{-0.0cm}
\mathbf{B}_{k, p} = \frac{\mathbf{\Phi}_{k}^{-1} \mathbf{h}_{k, p}}{\mathbf{h}_{k, p}^{\text{H}}\mathbf{\Phi}_{k}^{-1} \mathbf{h}_{k, p}},
\vspace{-0.0cm}
\end{equation}
where $\mathbf{h}$ and $\mathbf{\Phi}$ denote the steer vectors and noise correlation matrix, respectively. When $\mathbf{\Phi}$ is the identity matrix, the calculated beamformer corresponds to the DS case, and that of SD if $\mathbf{\Phi}$ is the diffuse noise correlation matrix. Here we uniformly sample the space with $P$ beams by adjusting the steer vector toward the target DOA. For example, if the circular array is employed and 36 basis beams are set, then we traverse the space from $0^{\circ}$ to $350^{\circ}$ with a step $\Delta\theta = 10^{\circ}$ to obtain the beam-space dictionary.

In the semi-learnable setting, we relax the trainable property of beam-space dictionary. Specifically, we set the parameters of $\mathbf{\Phi}$ to be trainable while steer vectors are fixed so that the noise correlation matrix can be optimized in the training process. Note that to guarantee the semi-positive property of the noise correlation matrix, its inversion is calculated by $\mathbf{\Phi}^{-1}=\mathbf{U}\mathbf{U}^{\text{H}}$~{\cite{zhang2022all}} and $\mathbf{U}$ is a lower triangular matrix.

For full-learnable scheme, a natural option is to let both the noise correlation matrix and steer vectors to be trainable but the dictionary is still calculated following the formula of Eq.~(\ref{eqn:eq14}). Besides, we investigate whether keeping the physical meaning of basis beam is necessary or not. After the initialization, the whole dictionary is switched to be trainable and it does not need to follow the formula shown in Eq.~(\ref{eqn:eq14}) in the training process.
\vspace{-0.25cm}
\subsection{Network Structure}
\label{sec:network-structure}
\vspace{-0.15cm}
The overall diagram of the proposed system is shown in Fig.~{\ref{fig:architecture}}. In general, any existing network structures can easily adapt to our framework, and in this study, we adopt the same structure as~{\cite{li2022taylorbeamformer}}. After the TI-BD, the noisy spectra from array are transformed into a beam set with $P$ beams, and we concatenate them along the channel dimension to obtain the tensor $\mathbf{Y}\in\mathbb{C}^{L\times K\times P}$. In the 0th-order module, a typical ``Encoder-Decoder'' structure is adopted with cascade squeezed temporal convolution modules (S-TCMs)~{\cite{li2021two}} in the bottleneck for sequence modeling. After that, sub-band LSTMs are utilized to estimate the activating matrix for beam mixing in each T-F bin. For the high-order terms estimation, we follow the recursive formula in Eq.~(\ref{eqn:eq13}) and multiple S-TCMs are adopted to model the distribution of the complicated derivative term. Finally, we superimpose both the 0th-order and high-order terms to obtain the target speech. Due to the space limit, interested readers may refer to~{\cite{li2022taylorbeamformer}} for more network details.
\vspace{-0.15cm}
\section{Experimental Setup}
\label{sec:experimental-setup}
\vspace{-0.15cm}
\subsection{Dataset Configuration}
\label{sec:dataset-configuration}
\vspace{-0.15cm}
We use the open-sourced LibriSpeech ASR corpus~{\cite{panayotov2015librispeech}} to synthesize the multi-channel noisy-clean pairs, where \textit{train-clean-100}, \textit{dev-clean}, and \textit{test-clean} are used for training, validation and testing, respectively. For directional noise source, we randomly select 20,000 types of noises from the DNS-Challenge noise set{\footnote{github.com/microsoft/DNS-Challenge/tree/master/datasets}}, whose duration is around 55 hours. We simulate multi-channel RIRs based on a circular array of seven microphones, where one microphone is placed in the center and the remaining six microphones are uniformly spaced on a circle. The radius of the circle is set to 4.25 cm. Without loss of generality, the microphone in the center is selected as reference. The room size ranges from 5-5-3 m to 10-10-4 m in the length-width-height format. The reverberation time (T$_{\text{60}}$) is sampled in the range of 0.1-1.0 s and the first 0.1 s of the room impulse response (RIR) with reverberation time shortening technique~{\cite{zhou2022single}} is convolved with clean speech to obtain the target speech. For each target speech, we randomly choose 1-3 positions to play the noise and the distance between the source and microphone center ranges from 0.5 m to 5.0 m. All the sources are assumed to be static without changing their positions within one utterance. The signal-to-noise ratio (SNR) is chosen from $[-5\rm{dB}, 10\rm{dB}]$. Totally, we generate 40,000 and 10,000 noisy-clean pairs for training and validation, respectively, and the average utterance length is around 4-second.

For model evaluations, two sets are set, namely Set-A and Set-B. In Set-A, we only set one directional noise with four DOA-difference cases, namely 0-15$^{\circ}$, 15$^{\circ}$-45$^{\circ}$, 45$^{\circ}$-90$^{\circ}$, and 90$^{\circ}$-180$^{\circ}$. For Set-B, 1-3 directional noises are placed with randomly selected DOAs. For both sets, around 50 noises from MUSAN corpus are selected~{\cite{snyder2015musan}}. Testing SNR ranges $[-5\rm{dB}, 5\rm{dB}]$, and 200 pairs are generated.
\vspace{-0.2cm}
\renewcommand\arraystretch{1.07}
\begin{table*}[t]
	\setcounter{table}{1}
	\caption{Quantitative comparisons with advanced baselines. The values are specified with PESQ/ESTOI/SI-SNR/DNSMOS formats.}
	\normalsize
	\setlength{\tabcolsep}{3pt}
	\centering
	\resizebox{0.85\textwidth}{!}{
		\begin{tabular}{c|c|cc|ccccc}
			\hline
			\multirow{2}*{Systems} &\multirow{2}*{Year}  &\multirow{1}*{Param.} &\multirow{1}*{MACs} &\multicolumn{4}{c}{Set-A  (DOA-difference between target speech and noise)} &\multirow{2}*{Set-B}\\
			\cline{5-8}
			& &(M) &(G/s) &0-15$^{\circ}$ &15$^{\circ}$-45$^{\circ}$ &45$^{\circ}$-90$^{\circ}$ &90$^{\circ}$-180$^{\circ}$ &\\
			\hline
			Noisy &- &- &- &1.73/45.39/-1.51/1.11 &1.73/44.34/-1.71/1.11 &1.71/45.03/-1.36/1.11 &1.66/44.89/-1.48/1.11 &1.63/41.13/-1.89/1.11\\
			TI-MVDR (Oracle) &- &- &- &2.54/75.32/9.08/1.91  &2.66/77.22/9.72/2.05 &2.71/78.26/9.63/2.05 &2.70/79.14/10.04/2.02 &2.46/73.05/8.78/1.82\\
			TI-MWF (Oracle) &- &- &-  &2.66/77.35/\textbf{12.15}/1.89 &2.74/78.98/\textbf{12.52}/2.01 &2.79/80.18/\textbf{12.85}/1.97 &2.77/80.96/\textbf{13.18}/1.97 &2.56/75.01/\textbf{10.79}/1.77\\
			FasNet-TAC(4ms) &2020 &2.65 &16.52  &2.53/70.33/8.44/2.43 &2.63/72.73/9.25/2.50 &2.75/75.51/10.01/2.52 &2.76/76.26/10.16/2.55 &2.59/71.51/8.69/2.40\\
			MMUB &2021 &\textbf{1.96} &8.35 &2.27/62.34/5.68/2.29 &2.26/62.11/5.81/2.29 &2.34/64.47/6.47/2.32 &2.26/63.31/6.25/2.32 &2.26/60.86/5.66/2.22 \\
			NSF &2022 &12.96 &4.99 &2.68/71.81/5.69/2.69 &2.69/71.69/5.99/2.71 &2.73/72.54/6.23/2.70 &2.69/72.15/6.35/2.75 &2.63/70.12/5.43/2.63 \\
			COSPA &2022 &3.66 &\textbf{1.16}  &2.27/62.22/5.72/2.10 &2.36/63.32/6.23/2.11 &2.51/67.13/7.05/2.19 &2.54/68.93/7.66/2.24 &2.31/61.07/5.29/2.08 \\
			FT-JNF &2022 &3.35 &54.36  &2.72/71.29/7.38/2.33 &2.84/74.59/8.06/2.44 &2.95/76.40/8.56/2.47 &2.97/77.36/8.89/2.52 &2.81/72.93/7.58/2.27 \\			
			EaBNet &2022 &2.82 &7.44  &3.00/78.87/8.79/2.60 &3.10/81.18/9.34/2.64 &3.21/82.93/10.05/2.64 &3.22/83.51/10.28/2.67 &3.04/79.69/8.82/2.56\\
			TayloyBF  &2022 &5.58 &8.62  &3.00/78.90/8.98/2.65 &3.12/81.49/9.69/2.71 &3.21/83.13/10.32/2.74 &3.23/83.64/10.57/2.76 &3.05/80.07/9.18/2.64\\
			\textbf{TaylorBM (Ours)} &2022 &5.63 &9.18 &\textbf{3.06}/\textbf{80.12}/9.59/\textbf{2.73} &\textbf{3.14}/\textbf{82.05}/10.14/\textbf{2.80} &\textbf{3.24}/\textbf{83.62}/10.75/\textbf{2.79} &\textbf{3.26}/\textbf{84.06}/10.86/\textbf{2.80} &\textbf{3.10}/\textbf{80.76}/9.62/\textbf{2.70}\\
			\hline
	\end{tabular}}
	\label{tbl:results-comparison}
	\vspace{-0.45cm}
\end{table*}
\renewcommand\arraystretch{0.88}
\begin{table}[t]
	\setcounter{table}{0}
	\caption{Ablation study on Set-B. ``F'', ``S'' and ``F'' respectively denote fixed, semi-learnable, and full-learnable type for beam dictionary. ``1'' means keeping the physical meaning of beam and that of not for ``2''. \textbf{BOLD} indicates the best score in each case.}
	\normalsize
	\centering
	\resizebox{0.99\columnwidth}{!}{
		\begin{tabular}{c|ccccccc}
			\specialrule{0.1em}{0.25pt}{0.25pt}
			\multirow{2}*{Entry} &Beam &\multirow{2}*{$P$} &Param. &MACs   &\multirow{2}*{PESQ$\uparrow$} &\multirow{2}*{ESTOI (\%)$\uparrow$} &\multirow{2}*{SI-SNR (dB)$\uparrow$}\\
			&Dic. & &(M) &(G/s) & & &\\
			\specialrule{0.1em}{0.25pt}{0.25pt}
			1a &F-DS &36 &5.63 &9.18 &2.97 &78.18 &8.65\\
			1b &F-SD &36 &5.63 &9.18  &2.97 &76.79 &8.21\\
			1c &S &36 &5.63 &9.17 &3.04 &79.86 &8.92\\
			1d &F1 &36 &5.63 &9.18 &3.05 &79.87 &9.12\\
			1e &F2 &36 &5.63 &9.18  &\textbf{3.10} &\textbf{80.76} &\textbf{9.62}\\
			\specialrule{0.1em}{0.25pt}{0.25pt}
			2a &F2 &3 &\textbf{5.54} &\textbf{8.43} &2.96 &77.29 &8.60\\
			2b &F2 &6 &5.55 &8.50 &3.06 &80.53 &9.42\\
			2c &F2 &12 &5.57 &8.64 &3.08 &80.53 &9.53\\
			2d &F2 &72 &5.73 &9.99 &\textbf{3.10} &80.75 &9.57\\
			\specialrule{0.1em}{0.25pt}{0.25pt}
	\end{tabular}}
	\label{tbl:ablation-studies}
	\vspace{-0.4cm}
\end{table}
\vspace{-0.15cm}
\subsection{Training configuration}
\label{sec:training-configuration}
\vspace{-0.15cm}
All the utterances are sampled at 16 kHz. 20 ms squared-root Hann window is selected with 50\% overlap between adjacent frames. 320 FFT is adopted, leading to 161 dimensions in the frequency axis. The power spectrum compression strategy is adopted to decrease the dynamic range and the compression factor is empirically set to 0.5~{\cite{li2021importance}}. Adam optimizer~{\cite{kingma2014adam}} is adopted, and 60 epochs are trained in total with the batch size of 6 at the utterance level. The learning rate is initialized at 5e-4 and will be halved if the loss value does not decrease for two epochs. Demo is available at \href{https://andong-li-speech.github.io/TaylorBM-Demo/}{\textcolor{red}{TaylorBM-Demo}}.
\vspace{-0.25cm}
\subsection{Comparison Benchmark}
\label{sec:comparison-benchmark}
\vspace{-0.15cm}
Following the empirical option in~{\cite{li2022taylorbeamformer}}, we set the Taylor order to three, \emph{i.e.}, $Q = 3$, as it well balances between trainable parameters and performance. For beam-space dictionary, the number of beam base $P$ is set to 36 and we also study the impact of different $P$ values in the ablation study. We compare with other advanced baselines, including MMUB~{\cite{9596418}}, NSF~{\cite{tan2022neural}}, FasNet-TAC~{\cite{luo2020end}}, COSPA~{\cite{halimeh2022complex}}, FT-JNF~{\cite{tesch2022insights}}, EaBNet~{\cite{li2022embedding}},  TaylorBF~{\cite{li2022taylorbeamformer}}, and oracle TI-MVDR, and TI-MWF.
\vspace{-0.40cm}
\section{RESULTS AND ANALYSIS}
\label{sec:results-and-analysis}
\vspace{-0.25cm}
\subsection{Ablation Study}
\label{ablation-study}
\vspace{-0.25cm}
We conduct the ablation study to analyze the impact of the beam-space dictionary in terms of type and number, whose metric results in terms of PESQ~{\cite{rix2001perceptual}}, ESTOI~{\cite{jensen2016algorithm}}, and SI-SNR~{\cite{le2019sdr}} are shown in Table~{\ref{tbl:ablation-studies}}. Several conclusions can be drawn. First, when the DS and SD are employed in the beam-space dictionary, we observe the worst metric scores. This is because FBs only exhibit decent characteristics in the specific noise fields. For example, DS beamformer yields the largest white array-gain while the SD beamformer has the largest directivity under the diffuse noise field. However, in practical acoustic scenarios, the noise field can be relatively complicated, and using a fixed beam-space dictionary may not be adequate to describe the spatial relations accurately. Then we switch the noise correlation matrix to be learnable, and from entry 1a(b) to 1c, one can see notable metric improvements. This reveals the significance to represent the time-invariant noise field properly. When both steer vectors and noise correlation matrix are learnable, only marginal performance gain is obtained. We attribute the reason as a dense spatial beam sampling strategy is adopted, \emph{e.g.}, 36, which is a complete spatial representation even with oracle steer vectors. Finally, we observe further improvements when the basis beam does not obey the physical formula of FB anymore, \emph{i.e.}, from entry 1d to 1e. This shows that current manually prior constraints may not be always necessarily the feasible option from the joint learning perspective~{\cite{zhang2021adl}}.

When $P$ increases from 3 to 36, consistent improvements are achieved, indicating that increasing the number of spatial bases can benefit the learning of spatial cue. However, when $P$ further increases to 72, no performance gain is obtained and thus 36 beam bases are employed hereafter.
\begin{figure}[t]
	\centering
	\centerline{\includegraphics[width=\columnwidth]{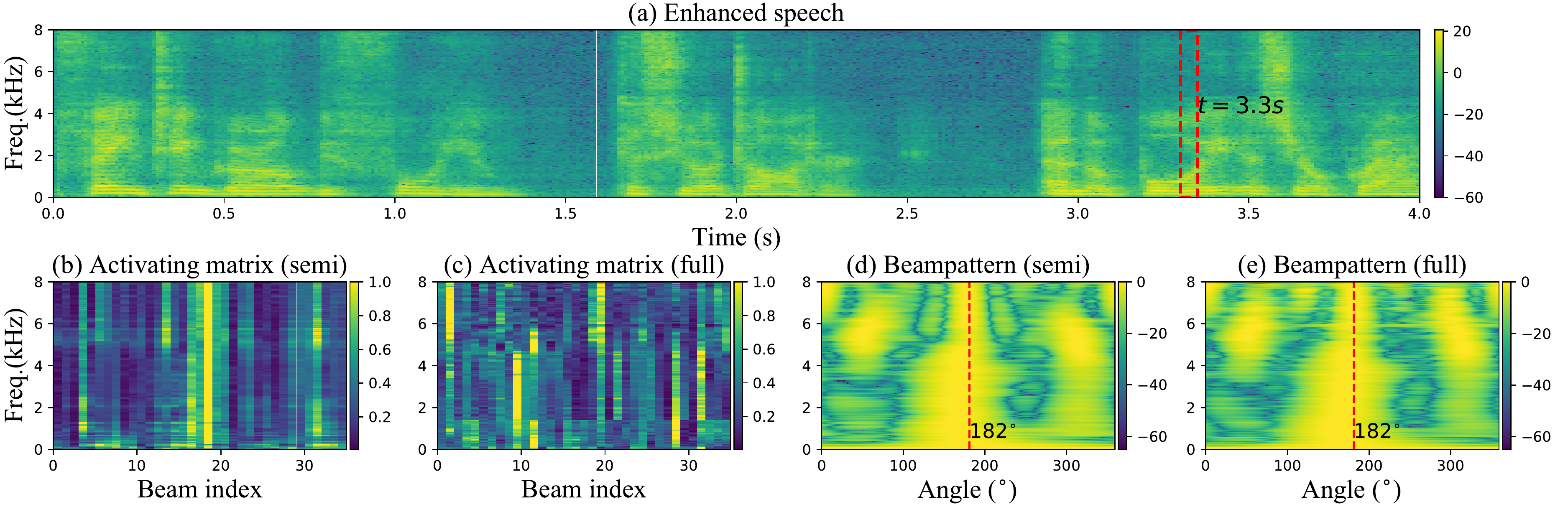}}
	\vspace{-0.15cm}
	\caption{An example visualization. (a) Enhanced speech by the proposed method. (b)-(c) Activation matrix for semi-learnable and full-learnable TI-BD, respectively. (d)-(e) Beampattern for semi-learnable and full-learnable TI-BD.}
	\label{fig:visualization}
	\vspace{-0.45cm}
\end{figure}
\vspace{-0.25cm}
\subsection{Results Comparison with Advanced Baselines} 
\label{results-comparison}
\vspace{-0.1cm}
Table~{\ref{tbl:results-comparison}} shows the quantitative comparisons with previous baselines. Besides PESQ/ESTOI/SI-SNR, DNSMOS~{\cite{reddy2021dnsmos}} is also adopted, which is an effective tool to simulate the subjective rating. One can see that overall, the proposed method yields better performance than the baselines. Compared with TaylorBF, we only add one differentiable layer with neglectable parameters, nonetheless, we observe notable performance improvements in multiple objective metrics. Also, as we convert the beamforming operation in the network into the beam-space dictionary learning and adaptive activating, the proposed method can exhibit better interpretability and provide more insight.

Fig.~{\ref{fig:visualization}} shows the visualization of the beam activation and beampattern at the time index $t$ = 3.3 s. The target source is located at 182$^{\circ}$ and three directional noises are placed at $\left\{1^{\circ}, 218^{\circ}, 237^{\circ}\right\}$. One can see that in Fig.~{\ref{fig:visualization}}(b), the beam response has a relatively large value in the beam index of 18 and low values in the 1st, 21-24th index, indicating that the propose model can properly select the beam component in the spatial sense. Interestingly, in Fig.~{\ref{fig:visualization}}(c), the activating distribution seems irregular and does not follow the expected spatial indication, and we attribute the reason as the beam-space dictionary and activating matrix are jointly learned and thus the basis beam may not obey the originally uniform spatial distribution. From Fig.~{\ref{fig:visualization}}(d)-(e), one can see that for either semi-learnable and full-learnable, the 0th-order term can effectively preserve the target source in the expected direction and suppress the noises by nulling, revealing that the 0th-order indeed serves as a spatial filter.
\vspace{-0.35cm}
\section{CONCLUSIONS}
\label{sec:conclusion}
\vspace{-0.20cm}
In this paper, we propose a Taylor-inspired all-neural beamformer dubbed TaylorBM for multi-channel speech enhancement . A beam-space dictionary is first employed to convert the received signals of different microphones into the dense beam distribution in the beam-space domain. Following Taylor's series expansion formula, in the 0th-order term, the spatial filter works by adaptively aggregating and mixing the beam components with various responses. And multiple high-order terms serve as the residual noise for post-processing. Experiments on a circular array with seven microphones reveal the superior performance of the proposed approach.   




\vfill\pagebreak
\bibliographystyle{IEEEbib}
\ninept

\end{document}